\documentclass[twocolumn, aps, preprintnumbers, amsmath, amssymb, superscriptaddress, nofootinbib]{revtex4}

\usepackage{graphicx,multirow}
\usepackage{longtable}
\usepackage{dcolumn}
\usepackage{bm}
\usepackage[all]{xy}
\usepackage{color}
\usepackage[usenames,dvipsnames]{xcolor}
\usepackage[unicode=true,pdfusetitle, bookmarks=true, bookmarksnumbered=false, bookmarksopen=false, citecolor=Turquoise, breaklinks=false, pdfborder={0 0 1}, backref=false, colorlinks=true, pdfpagemode=FullScreen]{hyperref}

\makeatletter

\newcommand{\fmslash}[2][0mu]{%
  \mathchoice
    {\fmsl@sh\displaystyle{#1}{#2}}%
    {\fmsl@sh\textstyle{#1}{#2}}%
    {\fmsl@sh\scriptstyle{#1}{#2}}%
    {\fmsl@sh\scriptscriptstyle{#1}{#2}}}
\newcommand{\fmsl@sh}[3]{%
  \m@th\ooalign{$\hfil#1\mkern#2/\hfil$\crcr$#1#3$}}
\makeatother

\newcommand{\lsim}{{\;\raise0.3ex\hbox{$<$\kern-0.75em\raise-1.1ex\hbox{$\sim$}}\;}}
\newcommand{\gsim}{{\;\raise0.3ex\hbox{$>$\kern-0.75em\raise-1.1ex\hbox{$\sim$}}\;}}

\newcommand{\beq}{\begin{equation}}
\newcommand{\eeq}{\end{equation}}
\newcommand{\bea}{\begin{eqnarray}}
\newcommand{\eea}{\end{eqnarray}}
\mathchardef\minus="002D

\usepackage{color}

\begin{document}
\title{Energy peak: back to the Galactic Center GeV gamma-ray excess}

\author{Doojin Kim\footnote{immworry@ufl.edu}}
\affiliation{Department of Physics, University of Florida, Gainesville, FL 32611, USA}
\author{Jong-Chul Park\footnote{jcpark@cnu.ac.kr}}
\affiliation{Department of Physics, Chungnam National University, Daejeon 305-764, Korea}
\preprint{CETUP2015-015}


\begin{abstract}
We propose a {\it novel} mechanism enabling us to have a {\it continuum} bump as a signature of gamma-ray excess in indirect detection experiments of dark matter (DM), postulating a {\it generic} dark sector having (at least) two DM candidates.
With the assumption of non-zero mass gap between the two DM candidates, the heavier one directly communicates to the partner of the lighter one.
Such a partner then decays into a lighter DM particle along with an ``axion-like'' particle (ALP) or dark ``pion'', which subsequently decays into a pair of photons, via a more-than-one step cascade decay process.
Since the cascade is initiated by the dark partner obtaining a non-trivial fixed boost factor, a continuum $\gamma$-ray energy spectrum naturally arises even with a particle directly decaying into two photons.
We apply the main idea to the energy spectrum of the GeV $\gamma$-rays from around the Galactic Center (GC), and find that the relevant observational data is well-reproduced by the theory expectation predicted by the proposed mechanism. Remarkably, the relevant energy spectrum has a {\it robust} peak at half the mass of the ALP or dark pion, as opposed to popular DM models directly annihilating to Standard Model particles where physical interpretations of the energy peak are {\it not} manifest.
Our data analysis reports substantially {\it improved} fits, compared to those annihilating DM models, and $\sim 900$ MeV mass of the ALP or dark pion.
\end{abstract}



\maketitle

\section{Introduction}

There is astrophysical and cosmological evidence that DM exists in the Universe (see, for example, Ref.~\cite{Bertone:2004pz}).
Relevant observations, mostly rooted in its gravitational effects, can be explained by postulating new stable particles, not belonging to particle species in the Standard Model (SM).
With this situation, there is a tremendous amount of effort to detect DM candidates: 1) direct detection experiments by measuring recoil energy of nuclei scattered off by DM, 2) indirect detection experiments by observing signals stemming from DM annihilation or decay, and 3) collider searches by actively producing DM particles and observing associated collider signatures.
Among those experimental efforts, satellite-based cosmic-ray detection experiments such as PAMELA~\cite{Adriani:2008zr,Adriani:2013uda}, AMS-02~\cite{Aguilar:2013qda,Accardo:2014lma}, and Fermi-LAT~\cite{Ackermann:2010ij,FermiLAT:2011ab} have received particular attention due to their great sensitivity to cosmic-ray signals, giving rise to better chance to have not only confirmation of the existence of DM but the information for deducing DM properties.

The Fermi-LAT collaboration has provided the public data based on their observations, and a $\gamma$-ray excess at $\mathcal{O}$(GeV) coming from the GC has been found.
In particular, it has recently reported in Ref.~\cite{TheFermi-LAT:2015kwa} that the excess exists, even assuming different foreground/background models.
The relevant program was initiated by Ref.~\cite{Goodenough:2009gk}, and their intriguing observation has been strengthened by a series of their follow-up analyses and other independent groups~\cite{Hooper:2010mq,Hooper:2011ti,Abazajian:2012pn,Hooper:2013rwa, Gordon:2013vta,Huang:2013pda,Abazajian:2014fta,Daylan:2014rsa,Lacroix:2014eea, Calore:2014xka,Calore:2014nla}.
Unlike other photon excesses such as 3.5 keV line~\cite{Bulbul:2014sua,Boyarsky:2014jta}, 511 keV line~\cite{Jean:2003ci}, and 130 GeV line~\cite{Bringmann:2012vr,Weniger:2012tx}, this is characterized by a {\it continuum} bump.
The basic claim is that the $\gamma$-ray excess spectrum is sufficiently consistent with the expected emission spectrum from charged particles in the SM into which DM particles are annihilated.
More specifically, the GeV excess is well-accommodated by a DM scenario where a pair of DM particles with a mass of $\sim30-40$ GeV annihilate into a $b\bar{b}$ pair with an annihilation cross section of $\langle\sigma v\rangle\sim 2\times 10^{-26} \hbox{cm}^3/\hbox{s}$~\cite{Daylan:2014rsa,Calore:2014nla}.
As an alternative annihilation channel, lepton pairs have been studied as well in Ref.~\cite{Lacroix:2014eea} where they pointed out the significance of the contributions coming from the diffuse photons from primary and secondary electrons that are produced in DM annihilation processes.
They further analyzed the data including the inverse Compton scattering and bremsstrahlung contributions from electrons, and found that the data is well-described by $\sim10$ GeV DM annihilating into a $\ell\bar{\ell}$ pair,
for which the associated annihilation cross section is given by $\langle\sigma v\rangle \approx (1-2)\times 10^{-26} \hbox{cm}^3/\hbox{s}$~\cite{Lacroix:2014eea}.
Moreover, Ref.~\cite{Agrawal:2014oha} showed that the GeV excess can be reproduced by other heavy SM final states such as $W^+W^-/ZZ/hh/t\bar{t}$ with a DM particle of $m_{\rm DM} \approx 80-200 ~{\rm GeV}$ and $\langle\sigma v\rangle \approx (2-8)\times 10^{-26} \hbox{cm}^3/\hbox{s}$,
depending on the final state
with the systematic uncertainties in the gamma-ray background modeling taken into account.
In Ref.~\cite{Calore:2014nla}, it was also shown that $gg/W^+W^-/ZZ/hh/t\bar{t}$ final states can provide a good fit to the excess with $m_{\rm DM} \approx 40-200 ~{\rm GeV}$ and $\langle\sigma v\rangle \approx (1-8)\times 10^{-26} \hbox{cm}^3/\hbox{s}$.

We remark that some of the realistic DM models have been proposed and studied:
for example, Refs.~\cite{Alvares:2012qv, Okada:2013bna, Modak:2013jya, Alves:2014yha, Ipek:2014gua, Basak:2014sza} for the $b\bar{b}$ final state through a Higgs portal type interaction,
Refs.~\cite{Kyae:2013qna,Kim:2015fpa} for $\ell\bar{\ell}$ final state,
Refs.~\cite{Boehm:2014bia, Ko:2014gha, Abdullah:2014lla} for DM annihilating to a pair of on-shell particles that subsequently decay into SM $f\bar{f}$ pairs,
Ref.~\cite{Elor:2015tva} even for $2^n$ pairs of SM $f\bar{f}$ final states from on-shell mediator pairs through multi-step cascades,
and Ref.~\cite{Kong:2014haa} for generic model constraints.
Although the recent report from the AMS-02 collaboration~\cite{AMS02} has started to rule out the $q\bar{q}$ final state dominant DM scenarios explaining the measured relic abundance~\cite{Giesen:2015ufa}, it is straightforward to invoke hybrid scenarios where $b\bar{b}$ and $\ell\bar{\ell}$ modes are mixed together.

One should notice that the astrophysical uncertainty in $\gamma$-rays coming from the GC in conjunction with the background modeling for the emission in the inner galaxy is still large.
In addition, pions from the collision between cosmic-rays and gas~\cite{Hooper:2010mq,Hooper:2011ti,Abazajian:2012pn,Gordon:2013vta} and millisecond pulsars~\cite{Hooper:2010mq,Hooper:2011ti,Abazajian:2012pn,
Gordon:2013vta, Abazajian:2014fta,Abazajian:2010zy} can be sources to the GeV scale $\gamma$-rays.
Therefore, they have been proposed as a different approach to interpret the excess although the relevant spectral shape appears too soft at the sub-GeV energy regime to accommodate the observed energy spectrum~\cite{Hooper:2013nhl}.
When it comes to the morphological feature for the observed excess, it is extended to more than $\sim10^{\circ}$ from the GC beyond the boundary of the central stellar cluster that could contain a large number of millisecond pulsars~\cite{Lacroix:2014eea}, and observed distributions of gas seem to give a poor fit to the spatial distribution of the signal~\cite{Lacroix:2014eea,Linden:2012iv,Macias:2013vya}.
We finally point out that very recently, another non-DM interpretation has been suggested by Ref.~\cite{Bartels:2015aea, Lee:2015fea}.
They basically came up with a new method to characterize unresolved point sources based on which the excess can be explained by a population of unresolved point sources, giving a distribution consistent with the observed GeV $\gamma$-ray excess in the relevant region.

\begin{figure}[t]
\centering
\includegraphics[width=6.0cm]{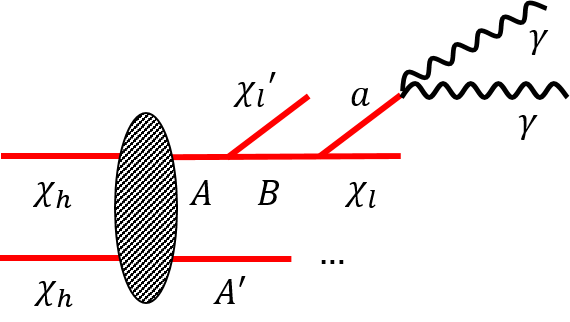}
\caption{\label{fig:model} The example ``dark'' cascade scenario under consideration.}
\end{figure}

Notwithstanding those potential issues, we here propose another {\it novel} mechanism to attain continuum energy spectra of the $\gamma$-ray excess, positing the DM interpretation.
We first remark that there could exist multiple DM species, and the DM models based on such a DM framework can give rise to not only non-trivial cosmological implications (e.g., ``{\it assisted freeze-out}''~\cite{Belanger:2011ww}) but interesting phenomenology (see, e.g., ``{\it boosted DM}''~\cite{Agashe:2014yua,Berger:2014sqa,Kong:2014mia}).
In this context, we assume a DM partner having a non-trivial and fixed boost, which could be achieved by the annihilation of another (heavier) DM.
The DM partner is further assumed to undergo a $\geq$ 2 step ``dark'' cascade decay, and emit, in the final step, a (lighter) DM particle along with an ALP or dark pion that subsequently decays into two photons.
FIG.~\ref{fig:model} schematically depicts the example ``dark'' cascade scenario of our interest.
We shall discuss the minimality of this DM scenario in conjunction with the elaboration of formalism later.
We emphasize that the relevant ALP or dark pion comes with a non-trivial boost distribution, typically rise-and-fall-shaped.
As a consequence, the resulting photon energy spectrum becomes a broad distribution.\footnote{It is well-known that sequential decays through and into SM particles typically invoke broadly-distributed energy spectra.
See, for example, Ref.~\cite{1971NASSP.249.....S} in the context of cosmic gamma-ray physics and Refs.~\cite{Agashe:2012bn,Agashe:2013eba} in the context of collider physics.}

A {\it novel} feature of this type of energy distribution is that the peak of the photon energy distribution is exactly the same as half the mass of the ALP or dark pion~\cite{1971NASSP.249.....S,Agashe:2012bn}.
In other words, such a peak position is {\it robustly} connected to a physical property. This is {\it not} the case for other interpretations such as the DM models directly annihilating to SM particles
because their energy peak highly depends on models of parton showering, diffusion mechanism, and so on from which the final photon spectrum is generated.
We emphasize that the proposed strategy is completely generic to be applicable to any continuum bump in cosmic-ray energy spectra, not restricted to the $\gamma$-ray excess at hand even if we employ it as a concrete and realistic example.

\section{Model set-up and formalism}

To set up the dark matter scenario to which our strategy is applied,\footnote{More dedicated dark matter model building to satisfy all features of the generic set-up is beyond the scope of this paper. We instead leave it as a future work~\cite{model}} we first introduce a dark sector containing (at least) two DM candidates.
We then assume that one of the DM particles is heavier than the other and the heavier one (henceforth denoted by $\chi_h$) communicates to the SM sector via the lighter one (henceforth denoted by $\chi_l$), i.e., relevant relic abundance can be evaluated by the assisted freeze-out~\cite{Belanger:2011ww}.
However, we further assume that $\chi_h$'s are {\it not} directly communicate to $\chi_l$'s, but through an intermediate state $A$ which subsequently decays into a $\chi_l$ and dark sector particle $a$ via, in general, multiple intermediate states.
For simplicity, a single intermediate state $B$ is taken throughout this letter, but we later discuss the necessity of it.\footnote{In general, $\chi_l$ may be either an unstable particle which subsequently decays into lighter particles, or even particle $a$. However, our argument does {\it not} depend on the attributes of $\chi_l$.
In addition, $A$ and $B$ can be generally either dark or SM sector particles, but we just assume that $A$ and $B$ are dark sector particles for simplicity.}
Particle $a$ eventually decays into a photon pair, being taken as a source of $\gamma$-ray excess.
In this sense, it could be regarded as an ALP and dark pion.
We again refer to FIG.~\ref{fig:model} demonstrating the example dark cascade scenario that we shall discuss in more detail.
Here $\chi_h$ pairs annihilate into $A$ plus $A'$ and $A$ decays into $B$ and a dark sector particle $\chi'_l$ solely for full generality.
Since subsequent dynamics of $A'$ and $\chi'_l$ is irrelevant to the later argument,\footnote{Particles $A'$ and $\chi'_l$ could even decay into SM particles unless they are severely constrained by other observational data (e.g., excess in cosmic-ray positron measurements,~\cite{Adriani:2008zr,Adriani:2013uda,Aguilar:2013qda,Accardo:2014lma}) or cosmological bounds.} we simply omit their further processes.

Throughout the later argument, we take the assumption that particles $A$ and $B$ are scalars or produced in an {\it un}polarized way unless specified otherwise.
With the assumption that $\chi_h$ is non-relativistic, it is straightforward to have the range of the boost factor of $B$, $\gamma_B$:
\bea
\gamma_B^-&\leq& \gamma_B \leq \gamma_B^+\,, \nonumber \\
\gamma_B^{\pm} &\equiv& \frac{E_B^*}{m_B}\gamma_A\pm\frac{P_B^*}{m_B}\sqrt{\gamma_A^2-1}\,, \label{eq:rangegammaa}
\eea
where $\gamma_A$ denotes the fixed boost factor of $A$ (i.e., $m_{\chi_h}/m_A$) and $E_B^*$ ($P_B^*$) denotes the energy (momentum) of $B$ measured in the rest frame of $A$.
As is well-known, the distribution in $\gamma_B$ is flat.
To develop the intuition on the boost of particle $a$, we cast its energy in terms of boost factor $\gamma_B$:
\bea
E_{a}=E_{a}^*\gamma_B+P_{a}^*\sqrt{\gamma_B^2-1}\cos\theta_{a}^*\,, \label{eq:Ea}
\eea
where $E_{a}^*$ ($P_{a}^*$) is the energy (momentum) of $a$ measured in the rest frame of $B$
while $E_a$ is the corresponding energy measured in the laboratory frame,
and $\theta_{a}^*$ is the angle of its emission measured from the boost direction of $B$.
Obviously, $\gamma_a$ becomes 1 at $\cos\theta_a^*=-1$ together with the relation of $\gamma_B=E_a^*/m_a$, which arises in much of the relevant parameter space.
The reason is that $\gamma_B$ itself is given by a range as in Eq.~(\ref{eq:rangegammaa}), not a fixed value like $\gamma_A$ so that such a special value of $\gamma_B$ can be easily covered.
It turns out that the boost distribution of $a$, $g(\gamma_a)$ typically develops a rising-and-falling shape with a plateau region in the middle of it as far as $\gamma_B^-$ is away from 1.
More details of $g(\gamma_a)$ are determined by the associated mass spectrum of particles and underlying dynamics such as spin correlation if allowed.
However, to avoid any unnecessary complexity later, we simply suppose that $g(\gamma_a)$ takes a rising-and-falling structure starting from $\gamma_a \approx 1$.

When it comes to photons emitted from $a$, the relevant $\gamma$-ray energy spectrum becomes more involved.
Now that the two photons are not in the rest frame of $a$, their energy should be Lorentz-transformed like Eq.~(\ref{eq:Ea}):
\bea
E_{\gamma}=E_{\gamma}^*(\gamma_a+\sqrt{\gamma_a^2-1}\cos\theta_{\gamma}^*)\,, \label{eq:Eg}
\eea
where
$E_{\gamma}^*$ ($E_{\gamma}$) is the photon energy measured in the rest frame of $a$ (in the laboratory frame)
and $\theta_{\gamma}^*$ is the angle of photon emission measured from the boost direction of $a$.
If $a$ is a scalar, $\cos\theta_{\gamma}^*$ becomes flat so that again the energy distribution in $E_{\gamma}$ becomes rectangular for a given $\gamma_a$.
Remarkably, $E_{\gamma}^*$ is commonly included in the rectangles for any $\gamma_a$~\cite{1971NASSP.249.....S,Agashe:2012bn}.
This implies that one simply ``stacks up'' such rectangles for all possible values of $\gamma_a$ to get the energy distribution of $\gamma$-rays, $f(E_{\gamma})$.
This procedure is then translated into performing a Lebesque-type integral over the $\gamma_a$'s contributing to the given $E_{\gamma}$~\cite{1971NASSP.249.....S,Agashe:2012bn}:
\bea
f(E_{\gamma})=\int^{\infty}_{\frac{1}{2}\left(\frac{E_{\gamma}}{E_{\gamma}^*}+\frac{E_{\gamma}^*}{E_{\gamma}} \right)}d\gamma_a\frac{g(\gamma_a)}{2E_{\gamma}^*\sqrt{\gamma_a^2-1}}\,. \label{eq:int}
\eea
However, in general, the above integral is {\it not} analytically doable, so we employ the asymptotic ansatz that was originally proposed in Ref.~\cite{Agashe:2012bn}.
\bea
f(E_\gamma)\sim \exp\left[-\frac{w}{2}\left(\frac{E_{\gamma}}{E_{\gamma}^*}+\frac{E_{\gamma}^*}{E_{\gamma}} \right)\right]\,, \label{eq:original}
\eea
where $w$ is the width parameter being responsible for the details of the distribution.
Its performance and applicability were very successful in a wide range of examples of collider signatures~\cite{Agashe:2012bn, Agashe:2013eba, Agashe:2012fs, Chen:2014oha, Agashe:2015wwa, Agashe:2015ike},\footnote{Similar idea was also used for non-standard interpretations of the 750 GeV diphoton excess at the LHC \cite{Cho:2015nxy}.}
and therefore we anticipate that it will be also relevant to the case at hand.
We are noticed that any ansatz should satisfy the symmetry property under the operation of $\frac{E_{\gamma}}{E_{\gamma}^*}\leftrightarrow\frac{E_{\gamma}^*}{E_{\gamma}}$, and thus we slightly modify the above ansatz by introducing another fit parameter $p$ determining the power of the argument in Eq.~(\ref{eq:original}):
\bea
f_M(E_\gamma)= N\exp\left[-\frac{w}{2}\left(\frac{E_{\gamma}}{E_{\gamma}^*}+\frac{E_{\gamma}^*}{E_{\gamma}} \right)^p\right], \label{eq:ansatz}
\eea
where $N$ is the overall normalization parameter.
We shall conduct data fit procedures later with this modified template.

An interesting observation should be mentioned here.
It is straightforward to see that $E_{\gamma}^*$ is the geometric mean of minimum $E_{\gamma}$ (i.e., $\cos\theta_{\gamma}^*=-1$) and maximum $E_{\gamma}$ (i.e., $\cos\theta_{\gamma}^*=+1$) for any $\gamma_a$ according to Eq.~(\ref{eq:Eg}).
This implies that the distribution in {\it logarithmic} $E_{\gamma}$ is symmetric with respect to $E_{\gamma}=E_{\gamma}^*$.
In other words, the center position of the photon energy distribution  appears as the unique peak in the energy distribution as mentioned before, and is exactly the same as $E_{\gamma}^*=m_a/2$, which can be used for mass measurement of particle $a$.
This robust link between the peak position and a physical property (the mass of $a$) is rather remarkable in contrast to other DM interpretations of the GeV $\gamma$-ray excess wherein such a connection is {\it not} manifest.
For more generic and detailed discussions and physical implications about these observations, see Ref.~\cite{Agashe:2012bn} and its follow-up works~\cite{Agashe:2013eba, Agashe:2012fs, Chen:2014oha, Agashe:2015wwa, Agashe:2015ike}.

Before moving on to our data analysis, we briefly discuss ``would-be'' minimal scenario where particle $A$ immediately decays into particle $a$, not through the intermediate state $B$.\footnote{See Ref.~\cite{Kim:2015gka} for a dedicated discussion on physical implications and features of this would-be minimal scenario with regard to the DM.}
Obviously, the boost distribution of particle $a$ in this case is given by a box shape (see Eq.~(\ref{eq:rangegammaa}) with $B$ replaced by $a$).
The resulting gamma-ray spectrum contains a plateau region in the middle of it unless the minimum $\gamma_a$ approaches 1.
We find that the GeV excess does not show any apparent plateau structure in the plane of $E_{\gamma}$ versus $dN/dE_{\gamma}$.
Of course, one could render $\gamma_a$ close to 1, but this demands an unnatural tuning among relevant mass parameters, i.e., the fixed $\gamma_A (=m_{\chi_h}/m_A)$ should be close to $E_a^*/m_a$.
Furthermore, the box-shaped probability distribution for $\gamma_a$ ensures a non-zero probability at $\gamma_a=1$.
This would result in a sharp spike at $E_{\gamma}=m_a/2$ in the $\gamma$-ray spectrum~\cite{Chen:2014oha}, which is {\it not} accommodated by the observational data as well.
We also explicitly checked whether the would-be minimal DM scenario can explain the GC $\gamma$-ray excess, and found that the relevant fit does not reproduce the data in a reasonable level.
Therefore, the would-be minimal scenario is highly disfavored based upon this series of observations.
Instead, the scenario delineated in FIG.~\ref{fig:model} can be considered minimal, while evading all the above unwanted situations.

\section{Data analysis and discussions}

\begin{figure}[t]
\centering
\includegraphics[width=8.4cm]{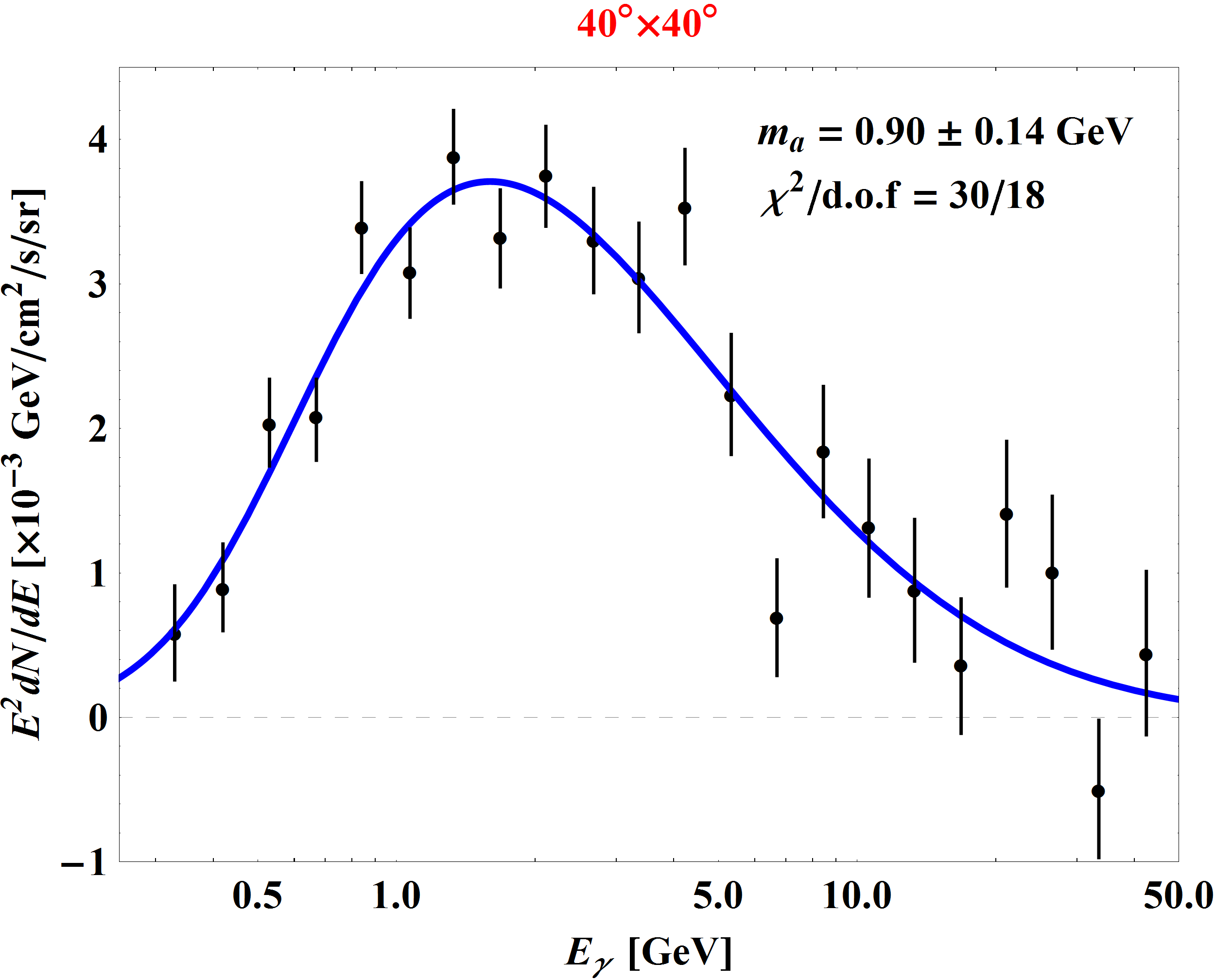} \\ \vspace{0.25cm}
\includegraphics[width=8.4cm]{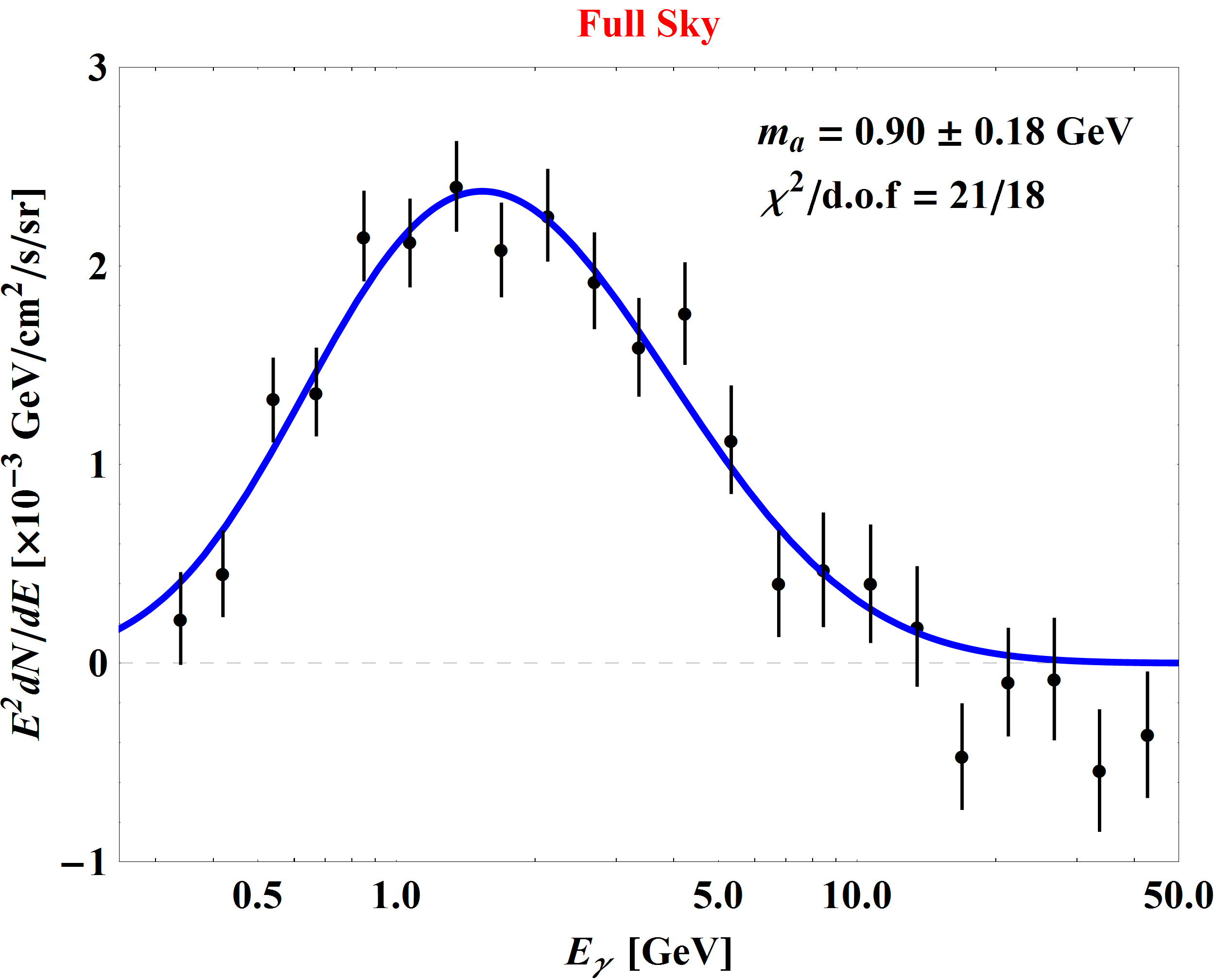}
\caption{\label{fig:fit} Upper panel: the $\gamma$-ray spectrum of the DM component taken from Ref.~\cite{Daylan:2014rsa}, corresponding to the data in ROI (i) ($1^{\circ}|b|<20^{\circ}$ and $|l|<20^{\circ}$).
The fit is performed with 22 data points, the best-fitted result is shown by the blue curve.
Lower panel: the same thing as in the upper panel, corresponding to the data in ROI (ii) (full-sky with $|b|>1^{\circ}$). }
\end{figure}

We next conduct the fit to the spectrum of the observed GC GeV $\gamma$-ray excess with the expected shape in Eq.~(\ref{eq:ansatz}).
The relevant data points are taken from Ref.~\cite{Daylan:2014rsa}.
Since their data is reported in the plane of $E_{\gamma}$ versus $E_{\gamma}^2\frac{dN}{dE_{\gamma}}$, i.e., weighted energy spectra, we practically perform the fit with $E_{\gamma}^2f_M(E_{\gamma})$.
The fits are done with both of their regions of interest (ROIs): (i) $1^{\circ}<|b|<20^{\circ}$ and $|l|<20^{\circ}$, and (ii) full-sky with $|b|>1^{\circ}$, where $b$ and $l$ are the Galactic latitude and longitude, respectively .
In FIG.~\ref{fig:fit}, our fit results are shown for both ROIs: the upper panel for ROI (i) and the lower panel for ROI (ii).
The data points are represented by black dots and their error bars are represented by black lines.
The best-fitted curves are exhibited by blue curves.

We see that our fits are in a rather good agreement with the observed $\gamma$-ray spectrum for both ROIs.
In other words, the data is well-reproduced by our fitting template.
To quantify the goodness of our fits, we evaluate the $\chi^2$ values which are 30 and 21 for ROI (i) and ROI (ii), respectively for 18 degrees of freedom (i.e., 22 data points subtracted by 4 fitting parameters such as $N$, $w$, $p$, and $E_{\gamma}^*$) between 0.3 and 50 GeV.
More quantitatively, our $\chi^2$ values are twice smaller than those from other studies such as Ref.~\cite{Daylan:2014rsa}.
These fit results indicate that our model scenario can provide a very good explanation for the observed $\gamma$-ray excess.
As advertised before, the fit can tell us about the mass of the particle decaying into two photons, $m_a$.
The best-fitted $m_a$'s extracted from $E_{\gamma}^*$ values for both ROIs are given by
\bea
m_{a}=\left\{
\begin{array}{l l}
0.90 \pm 0.14 \hbox{ GeV} & \hbox{ for ROI (i)} \\ [1mm]
0.90 \pm 0.18 \hbox{ GeV} & \hbox{ for ROI (ii)},
\end{array}\right.
\eea
where the reported errors here are $1\sigma$ statistical uncertainty.
Note that the peak positions in the distributions in FIG.~\ref{fig:fit} are slightly away from $m_a/2$ unlike the previous theoretic argument because they are {\it weighted} energy spectra so that the peak position becomes shifted to the higher energy region.
The fitted masses for both ROIs are in a great agreement to each other, more convincing ourselves that the data in both signal regions comes from a common source.

\section{Conclusions}

In conclusion, we have shown that for the GC GeV $\gamma$-ray excess, an alternative avenue of generating {\it continuum} bump is available with the DM interpretation.
More specifically speaking, as a basic and minimal set-up, we introduced a dark sector in which there exist (at least) two DM candidates, the heavier and the lighter ones.
The heavier DM particle is set to communicate to the SM sector via the lighter DM particle, but heavier DM particles are assumed to annihilate into an intermediate state rather than directly into lighter DM particles.
This intermediate state further decays into a lighter DM particle together with an ALP or dark pion, which directly decays into a pair of photons, via intermediate dark sector state(s).
In this scenario, we showed that the first intermediate state comes with a fixed boost factor, and in turn, the second intermediate state comes with a flat boost distribution. Thus, the ALP or dark pion gets varying boost factors characterized by a rising-and-falling structure, naturally resulting in a continuum photon energy spectrum even though it decays directly into two photons.

One of the {\it novel} features in this DM interpretation is that the relevant energy spectrum has a symmetry property with respect to the rest-frame energy of photons ($E_{\gamma}^*$), which motivates and constrains the form of the ansatz describing the relevant energy distribution.
Therefore, the mass of the particle decaying into a photon pair can be measured by reading off $E_{\gamma}^*$, which is given by half the mass of the ALP or dark pion, from the fit procedure.
In addition, considering the DM scenario to be employed here, one could take the GeV excess as a ``smoking-gun'' signal for the non-minimal dark sector, which currently receives increasing attentions.

With the given model set-up, we have introduced a template for the GC GeV $\gamma$-ray excess.
We then have studied how well such a template can reproduce the observed energy spectrum of GC $\gamma$-rays.
Performing the standard $\chi^2$ fitting procedure, we have obtained $\chi^2=$ 30 and 21 for ROI (i) and (ii), respectively, and both ROIs have consistently reported that the mass of the ALP or dark pion is $\sim900$ MeV.
We expect that these values will be better-established as the Fermi-LAT collaboration accumulates more cosmic-ray data.

In forthcoming work~\cite{model}, we shall construct realistic DM models to accommodate the basic set-up that we discussed in this letter.
In particular, we shall investigate phenomenological and cosmological implications of the models in the aspects of direct detection, collider phenomena, and dark matter relic density.
In addition, we shall study the connection between underlying DM model parameters and the fit parameters other than $E_{\gamma}^*$ in the context of a concrete DM model.

\section*{Acknowledgments}

We would like to thank Kaustubh Agashe and Kyoungchul Kong for a careful reading of the draft and useful discussions.
D.K. is supported by the LHC Theory Initiative postdoctoral fellowship (NSF Grant No. PHY-0969510) and
J-C.P. is supported by Basic Science Research Program through the National Research Foundation of Korea funded by the Ministry of Education (NRF-2013R1A1A2061561).
We appreciate CETUP* (Center for Theoretical Underground Physics and Related Areas) for its hospitality during conceiving this work.

\end{document}